# Chapter 5
**Nanomaterials From Imogolite: Structure, Properties and Functional Materials**

Erwan Paineau,* Pascale Launois*
Laboratoire de Physique des Solides, UMR CNRS 8502, Univ. Paris Sud, Université Paris Saclay, 91405 Orsay Cedex, France
Corresponding authors: E. Paineau (Email: erwan-nicolas.paineau@u-psud.fr); P. Launois (Email: pascale.launois@u-psud.fr)

**Abstract**
Hollow cylinders with a diameter in the nanometer range are carving out prime positions in nanoscience. Thanks to their physico-chemical properties, they could be key elements for next-generation nanofluidics devices, for selective molecular sieving, energy conversion or as catalytic nanoreactors. Several difficult problems such as fine diameter and interface control are solved for imogolite nanotubes. This chapter will present an overview of this unique class of clay nanotubes, from their geological occurrence to their synthesis and their applications. In particular, emphasis will be put on providing an up-to-date description of their structure and properties, their synthesis and the strategies developed to modify their interfaces in a controlled manner. Developments on their applications, in particular for polymer/imogolite nanotubes composites, molecular confinement or catalysis, are presented.

## 5.1. Occurrence and Distribution of Imogolite

Imogolite is a naturally occurring aluminosilicate nanotube, first described in 1962 by Yoshinaga and Aomine, in the Kuma basin in the Kumamoto Prefecture (Japan) [1]. The name of this unknown fibrous mineral has been proposed in reference to the parent material "imogo", a brownish yellow volcanic ash soil [2,3] (Figure 5.1a-c). The Commission on New Minerals and Mineral Names of the International Mineralogical Association first concluded that it did not warrant classification as a distinct mineral type [4]. The name was definitely approved by the International Association for Clay Studies in 1971 [5], with the recognition of imogolite as a new mineral species.

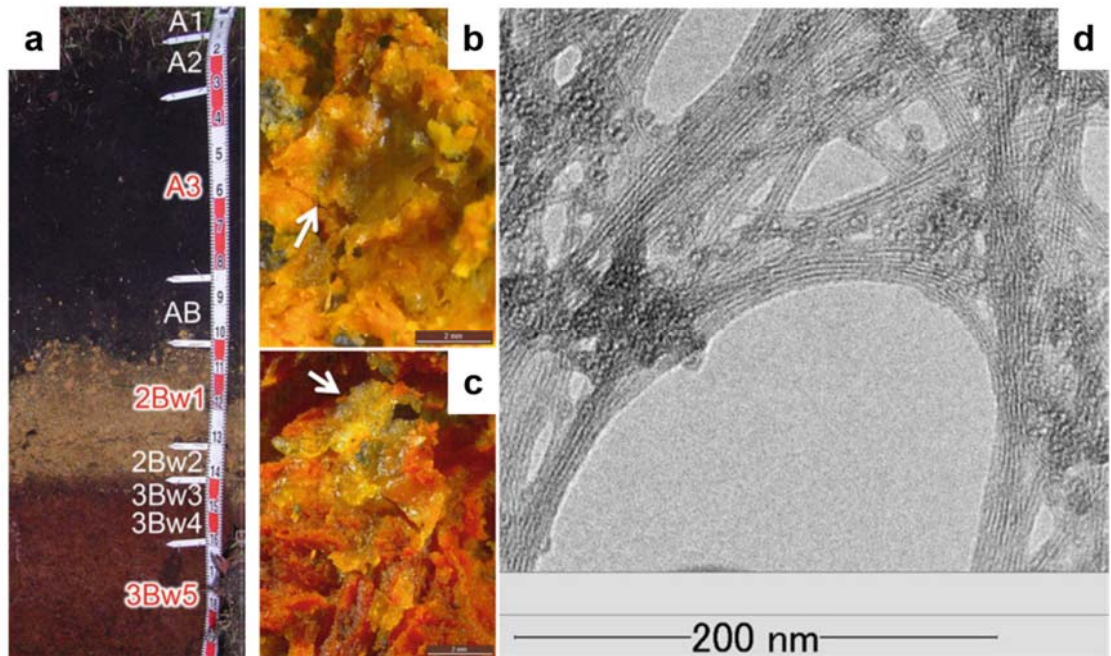

**Figure 5.1.** (a) Soil profile from Kiwadashima, Tochigi, Japan, (b-c) Optical observations of gel films (white arrows) found in the 2Bw1 and 3Bw5 horizons, respectively, (d) Typical transmission electron microscopy (TEM) micrograph of imogolite extracted from the gel-like film, in association with allophane (spherical particles). Adapted with permission from [3].

Since its first description in 1962, imogolite was found to occur worldwide either in almost pure form as macroscopic gel-like films in weathered pumice beds [6–9] or in B horizons of allophanic Andosols as a typical weathering product of pyroclastic materials under a humid climate and well-drained conditions [10–16]. However, the occurrence of natural imogolite is not restricted to volcanic ash soils since it was later recognized in podzolized horizons of Spodosol, formed from non-volcanic parent materials in colder regions of the world [17–21]. Although the mechanisms of imogolite formation in natural environment are not fully elucidated, it is well admitted that the presence of volcanic ash in these soils may enhance their formation since the dissolution of volcanic glass provides a readily available source of soluble silicic acid ($H_4SiO_4$) [19]. Beyond Earth deposits, imogolite was recently found in the layered outcrops at Mawrth Vallis in Mars, suggesting a period of rapid weathering and it is believed that it could be used as a marker in the evolution of the Martian climate [22,23].

Prior to use, natural imogolite needs to be extracted from parent material, which usually contains organic and inorganic impurities, as schematically described in Figure 5.2 [24].

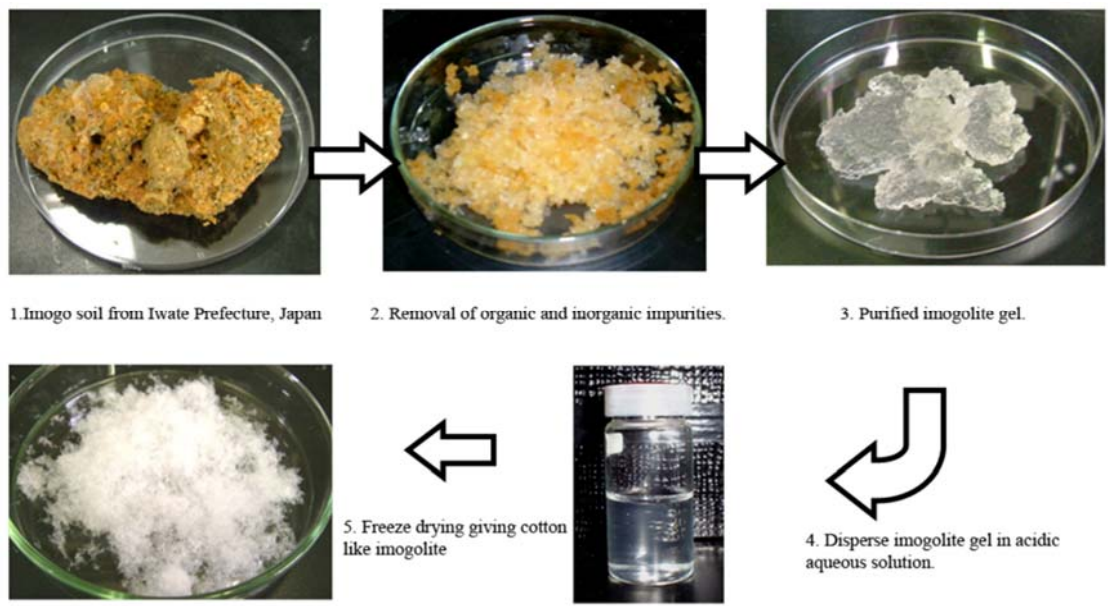

**Figure 5.2.** Schematic representation of the different purification steps of natural imogolite. Reproduced with permissions from [24] © 2012 Ma et al; licensee Beilstein-Institut.

Typically, soils samples are treated with an aqueous solution of hydrogen peroxide to remove organic contaminants, followed by citrate-bicarbonate-dithionite treatment to dissolve extractable oxides. Purified imogolite gels are then washed by centrifugation and redispersed in weak acidic solution. Finally, air-drying or freeze-drying processes are often applied to obtain a cotton-like substance for further characterization but packing of nanotubes in bundles may occur during such processes, which may impede their full dissociation afterwards.

Transmission Electron Microscopy (TEM) observations remain the most common approach to identify imogolite minerals in soil samples. Typical TEM micrograph is presented in Figure 5.1d, showing the characteristic thread-like morphology of imogolite bundles, each thread consisting of bundled "fibre" units with diameter between 2 and 3 nm [8,25,26]. In 1970, Wada et al. [27] were the first to propose, thanks to improvements in TEM experiments, that the parallel lines observed in TEM micrographs represent the walls of a hollow tube rather than a fibre structure.

## 5.2. Structure and Physico-Chemical Properties of Imogolite
### 5.2.1. Structure of imogolite

During the ten years following the observation of fine fibrous imogolite particles in soils, their structure remained unclear. Due to the presence of broad modulations in X-ray scattering diagrams, imogolite was claimed to be a non-crystalline or poorly crystalline mineral. Different structural models were proposed based on X-ray scattering and electron diffraction [26,28,29], involving paracrystalline order as well as various chains of $AlO_6$ octahedra linked sideways by $SiO_4$ tetrahedra. As mentioned above, in 1970, Wada et al. [27] showed, on the basis of high resolution electron micrographs, that the unit fibre is a hollow tube, opening new perspectives in the determination of the imogolite structure.

The structure of imogolite nanotubes (INT) was solved in 1972, in an article which brought together the different teams previously involved in its structural analysis [30]. Its nominal composition is $(OH)_3Al_2O_3Si(OH)$, where hydroxyls groups and atoms are labelled here from outside to inside the nanotube. Imogolite nanotube consists of a curved di-octahedral gibbsite-like layer $(Al(OH)_3)$ with isolated $(SiO_3)OH$ tetrahedron units connected via covalent bonding between three mutual oxygen atoms, as it is shown in Figure 5.3a. The curvature effect can be understood in a ultra simplified way by noting that the distance between adjacent oxygen atoms on the tetrahedron is shorter than that on the planar gibbsite-like sheet. Rolling would thus allow adjusting the distances. The gibbsite-like sheet is rolled in such a way that aluminium octahedra have a zig-zag arrangement perpendicularly to the nanotube axis, as it is highlighted in Figure 5.3a. The nanotube structure is characterized by the number *N* of octahedra along such a zig-zag circumference. In their seminal article on the structure of natural imogolites [30], Cradwick and co-workers could not determine the exact value of *N* between $N$ = 10, 11 and 12. It has not yet been done unambiguously since then, despite intensive research. The diameter of imogolite nanotubes obtained artificially (see section 5.3) is slightly larger than that of natural imogolite nanotubes, indicating larger *N* values.

As already mentioned, X-ray scattering diagrams of imogolite nanotube samples consist of a limited number of broad modulations. It is not due to a poor crystallinity of imogolite, as it was first assumed, but to the nanometric lateral extend of the nanotubes. A methodogy was recently proposed to determine their atomic structure determination from wide-angle X-ray scattering experiments [31]. Monet et al. applied the methodology to methylated imogolite nanotubes, where methyl groups replace the inner hydroxyl groups (see section 5.3). They evidenced a different rolling mode for these modified imogolite nanotubes, with an 'armchair' arrangement of the aluminium octahedra along the tube circumference, as it is illustrated in Figure 5.3b [31].

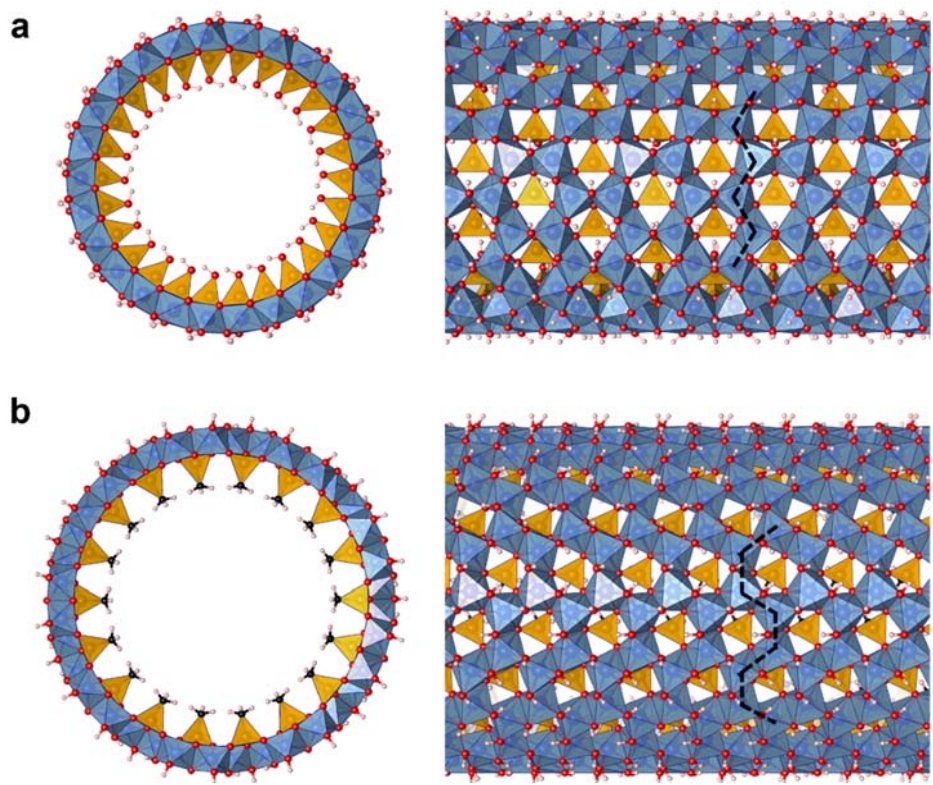

**Figure 5.3.** (a) Schematic top and side view of a *N*=13 imogolite nanotube. Ochre, blue, red and white atoms are Si, Al, O and H atoms, respectively. Ochre tetrahedra (blue octahedrons) are $SiO_4$ tetrahedra ($AlO_6$ octahedrons), with O atoms at the top and Si (Al) atom at the center. The dashed black line on the right view illustrates the zig-zag configuration. (b) Top and side view of a methylated imogolite nanotube. Dark atoms are C atoms and ochre tetrahedra are now $SiO_3C$ tetrahedra. The dashed black line on the right view illustrates the armchair configuration.

The family of imogolite nanotubes can be extended to methylated imogolite nanotubes or to germanium-based imogolite nanotubes, where silicon is replaced by germanium (see Section 5.3). With diameters of the order of the nanometer, INTs can be considered as true inorganic analogs of the well-known carbon nanotubes. But contrary to single-walled carbon nanotubes or other inorganic single-walled nanotubes such as BN or $MoS_2$ ones [32,33], whose strain energy decreases monotonically with increasing diameter, INTs present a well-defined energy minimum as a function of their diameter or of their chirality [31,34–39].

It is worth concluding this paragraph by emphasizing the singularity of imogolite compared to other clays. Let's briefly compare it here to halloysite presented in Figure 5.4 (for a review article about halloysite, see e.g. [40]). Halloysite is generally polydisperse and does not form single-walled nanotube like imogolite. The wall of an halloysite multi-walled nanotube (or nanoscroll, see Figure 5.4b) has nominal formula $Al_2Si_2O_5(OH)_4$. It consists in a gibbsite-like sheet with which $SiO_4$ tetrahedra share only one oxygen atom (Figure 5.4a). In other words, tetrahedra are head to toe in imogolite and in halloysite. Tetrahedron orientation in imogolite is exceptional for a clay mineral. The external surface of halloysite is covered by polymerized $SiO_4$ tetrahedra forming a true layer [41], while it is the internal surface of imogolite, which consists in a silicon-based sheet. It is covered by isolated $SiO_3(OH)$ tetrahedra, a unique feature for a clay.

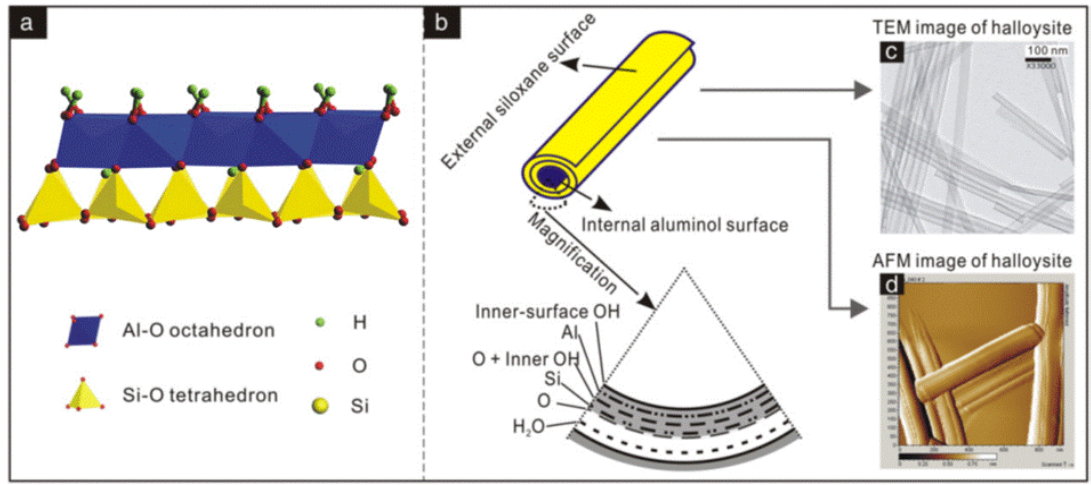

**Figure 5.4.** Schematic representation of (a) the wall structure of halloysite and (b) its tubular morphology. Typical (c) transmission electron microscopy (TEM) and (d) atomic force microscopy (AFM) images of halloysite nanotubes. Adapted with permissions from [41].

**5.2.2. Physico-chemical Properties of Imogolite**

The structure of imogolite is at the origin of its peculiar properties. Probably the most significant behavior of imogolite nanotubes is their monodispersity in diameter and chirality [30,31]. It is a major advantage for the detailed study of their physico-chemical properties or the development of some targeted applications (see section 5.4). Concerning their electronic properties, Bursill et al. were the first to suggest a large electronic band gap structure for imogolite [42]. According to different simulations, aluminosilicate nanotubes are insulators with a calculated band gaps higher than 4 eV [36,38,43,44]. Several efforts have therefore been prompted recently to reduce the band gap, in particular by doping the imogolite structure via isomorphic substitution [45–47] (see section 5.3).

Like other clay minerals, imogolite is subject to dehydroxylation and high temperature structural transformations. Its thermal stability is relatively low with dehydroxylation occurring at temperatures above T > 300°C [1,8,48–50]. Dehydroxylation process actually preludes to subsequent structural collapse. Further heating treatment led either to the occurrence of an X-ray amorphous phase [51,52] or to a lamellar phase [49,53], corresponding to the partial breakdown of the nanotube structure. At even higher temperatures, thermogravimetric analysis (TGA) of INTs exhibits an exothermic peak above 950°C. It was first related to the formation of $\gamma$-alumina [1]. It seems however to correspond to a mullite crystalline phase ($3Al_2O_3 \cdot 2SiO_2$) [52].

As far as the imogolite stability is concerned, knowledge of its mechanical properties is of prime importance towards future developments in material science. Several computational studies have reported that the Young modulus of INTs, characterizing the stiffness of the nanotube, is in the range 100-300 GPa [36,38,54,54], i.e. it is of the same order of magnitude as for halloysite (140 GPa) [55], $MoS_2$ nanotubes (150 GPa) [56] or chrysotile (160-280 GPa) [57]. However, experimental studies of INT mechanical properties are still missing. Moreover, the possible existence of various types of defects in the imogolite wall [58] may strongly modify the mechanical properties of the nanotubes [59,60]. There are several indications that imogolite nanotubes can easily deform, more specifically when they are assembled in bundles. Molecular dynamic calculations performed on aluminosilicate imogolites show that the basis of interacting nanotubes can deform to become oval [34,61]. Recent numerical calculations [59] also indicate that partial dehydroxylation of the inner surface silanols of the aluminosilicate nanotube should induce substantial deformations of the nanotubes. The authors concluded that the control of the degree of dehydroxylation by heat treatment may allow one to tune the electronic and mechanical properties of imogolite nanotubes [59]. From an experimental point of view, Amara et al. [62] gave the first evidence of a deformation, namely hexagonalization, of imogolite nanotubes at room temperature, when they are assembled in bundles, thanks to a systematic method developed to analyze X-ray scattering diagrams as a function of the nanotube shape.

It was early recognized that imogolite nanotubes present interesting colloidal properties. In their seminal paper, Yoshinaga and Aomine evidenced that INTs aggregate in concentrated NaCl solutions with a jelly-like appearance, while air-dried INTs recover their swelling properties in weak acidic suspensions (pH < 4) [1]. It is now well proved that imogolite can form transparent gels, which generally fill instantaneously the whole suspension volume by fast coagulation mechanisms [63]. Furthermore, suspensions of imogolite nanotubes form a true liquid-crystalline nematic phase at very low volume fractions, either for natural [64–67] or synthetic imogolite nanotubes [68,69], as expected for highly elongated colloidal nanorods. Moreover, Paineau et al. discovered that, at slightly higher imogolite concentration, INTs also form a liquid-crystalline hexagonal columnar phase (Figure 5.5a), which is much more stable to dilution than reported so far in literature [69]. Thanks to its low visco-elasticity, this columnar phase is easily aligned in an alternating electric field, in contrast with usual columnar liquid-crystal phases [70,71] (Figure 5.5b). This significant advantage could be exploited for the elaboration of anisotropic nanocomposite materials (see section 5.4).

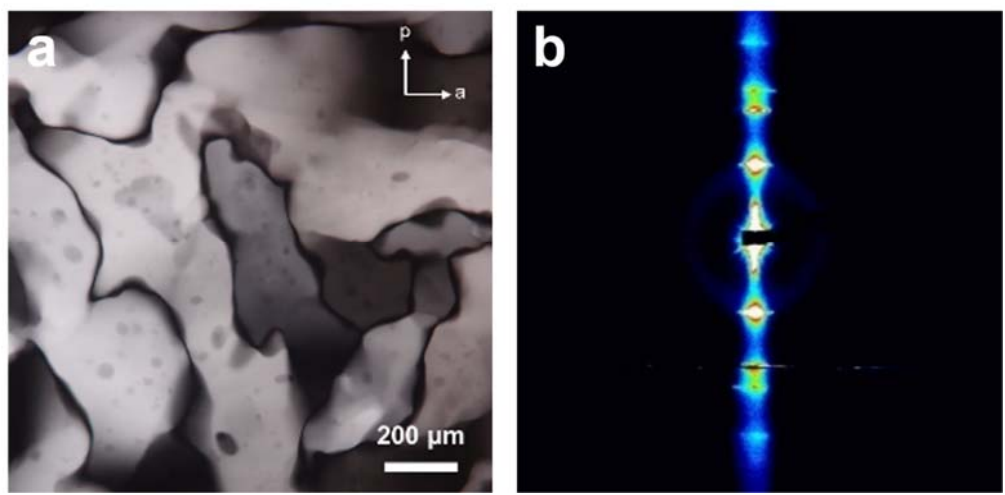

**Figure 5.5.** (a) Optical observation between crossed polarizers (white arrows) of a hexagonal columnar phase (Col$_H$) in aqueous suspensions of aluminogermanate imogolite nanotubes (volume fraction = 0.22%). (b) Two-dimensional small-angle X-ray scattering image of the Col$_H$ phase aligned under electric field. Adapted with permissions from [69].

Furthermore, imogolite nanotubes present high capacity to bind anions as originally evidenced in geological context [72–75]. Electrophoretic measurements reveal that the outer-surface charge of imogolite is positive over a wide range of pH, with point of zero charge (PZC) values higher than 7 [73,76–78], which can be used to control nanotubes bundling by anion condensation. Recent X-ray scattering measurements confirmed this assumption [79]. It was shown that, even at low ionic strength, imogolite nanotubes tend to organize in small bundles after solvent evaporation. Altogether, experimental observations on liquid crystal phases [69] as well as those on nanotube bundling [79] strongly suggest the dominant role of electrostatic interactions between imogolite nanotubes. Gustafsson proposed that the structural charge may arise from elongations of Al-O bonds or shortening of Si-O ones, inducing a weak positive (negative) charge on the outer (inner) surface of imogolite [80], an hypothesis confirmed with density functional theory (DFT) simulations [36,38,39,44].

### 5.3. Synthesis and Modification of Imogolite
### 5.3.1. Synthesis of Imogolite Nanotubes

From an industrial point of view, the main drawback of natural imogolite is that they do not form large deposits compared to other natural one-dimensional clay minerals, such as halloysite, sepiolite or palygorskyte [40,81,82]. But their synthesis was developed quite quickly after their discovery. Farmer et al. [83] proposed in 1977 a sol-gel method to achieve synthesis of INTs in aqueous suspension and under mild conditions around 100°C. Typically, the synthesis of INTs can be divided in four steps as summarized in Figure 5.6.

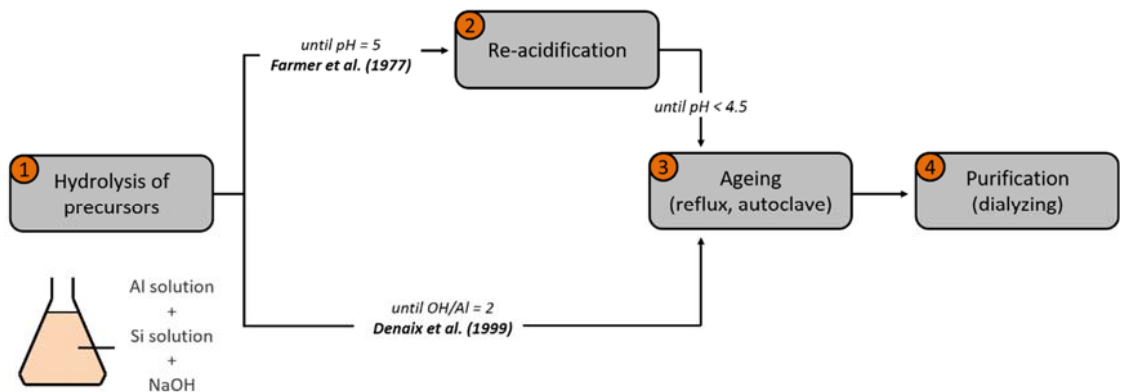

**Figure 5.6.** Flowchart for the two main synthesis methods of imogolite nanotubes. From Farmer et al. in [83] and Denaix et al. in [84].

The first one consists in the hydrolysis of Al and Si precursors by the addition of a NaOH solution under vigorous stirring. In a second step, the pH is adjusted to 4.5 by addition of a mixture of HCl and acetic acid promoting condensation of Al and formation of roof-tiled imogolite precursors, also referenced as proto-imogolite [85–89]. Let's note here that Denaix et al. showed that this step can be skipped by controlling the NaOH concentration up to an hydrolysis ratio [OH]/[Al] = 2 [84]. The third step consists in ageing the solution under hydrothermal conditions either by reflux or in autoclave to allow the growth of nanotubes. After sufficient time, the product is recovered at room temperature and purified by washing against water (step 4) (Figure 5.6). It should be underlined that compared to carbon nanotubes synthesis, the synthesis of imogolite is a low temperature process.

Since Farmer et al. article [83], several improvements have been proposed to increase the purity and the yield of the synthetic product by changing either the nature of the precursors, their concentration or their stoichiometric ratios [90–96]. Furthermore, several authors reported that the formation of imogolite nanotubes might depend on the step durations [83,97,98], by changing the temperature of the synthesis [99,100] or by modifying the method of heating [101,102].

Imogolite synthesis is commonly performed with Al alkoxide (ASB: aluminum tri-sec-butoxide), $AlCl_3$ or $Al(ClO_4)_3$ precursors, the latter offering a lower complexity ability, hence reducing the formation of by-products [103]. Although different Si sources have been explored [83,92,104,105], the use of alkoxide such as tetraethoxysilane (TEOS) is recognized to promote imogolite growth [51] due to its slow condensation kinetics. In addition, it was shown that a slight excess of Si precursor could inhibit the formation of aluminium hydroxides such as gibbsite or boehmite [90,106–108]. Moreover, several authors suggested that higher concentration of precursors or the nature of the counter-ions could prevent tube formation while promoting spherical-like nanostructures such as allophanes [96,108–111]. The synthesis of INTs thus requires careful adjustments.

### 5.3.2. Modification of Imogolite Nanotubes

An advantage of INTs over other nanotubes is that one can tune its morphology and wall chemistry by straightforward approaches. Yucelen et al. [88] suggested that the diameter of imogolite nanotubes can be changed by controlling the shape of nanometer-scale precursors formed in the early stage of the synthesis by changing the nature of the protic acid used (HCl, HClO$_4$, CH$_3$COOH) (Figure 5.7a).

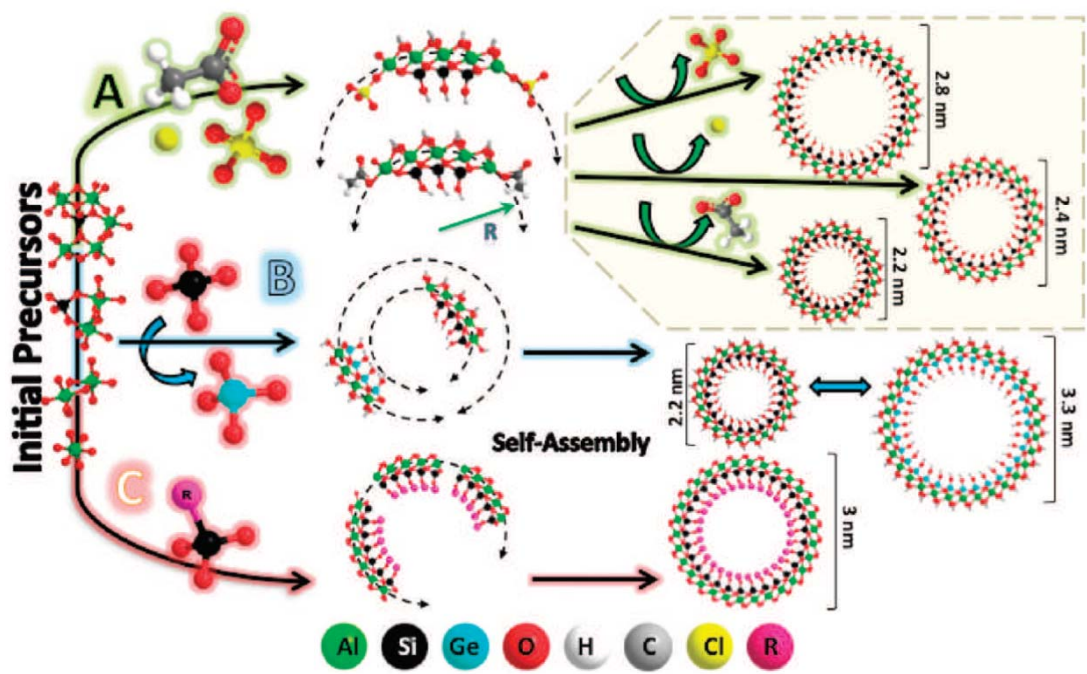

**Figure 5.7.** Schematic representation of shape control of imogolite nanotubes. (A) Anion complexation during the hydrolysis step; (B) isomorphic substitution; (C) inner functionalization by using specific organosilicate (or germane) tetrahedra. Reprinted with permissions from [88] © 2012 American Chemical Society.

One of the most blatant progress in INT synthesis is probably the replacement of silicon by germanium through isomorphic substitution (Figure 5.7b). Initially explored by Wada [112], the synthesis of aluminogermanate INTs (Ge-INT) has mainly risen in recent years [106,107,113]. Ge-analogues are produced in large quantities from decimolar initial concentrations [113]. Furthermore, Maillet et al. [114,115] evidenced that Ge-INTs could exist not only as single-walled (SWINT) but also as double-walled imogolite nanotubes (DWINT) depending of the initial concentration of aluminum precursor [116]. The produced Ge-based imogolite nanotubes were shorter than Si-based ones but Amara et al. [117] recently reported the synthesis of micron-long germanium-based imogolite nanotubes by producing the hydroxyl ions by thermal decomposition of urea (CO(NH$_2$)$_2$), bypassing the process of slow injection of NaOH. The inner diameter increases from ~ 1.5 to 3 nm from Si to Ge single-walled INTs. SWINT diameter can be tuned finely by adjusting the concentrations in initial Si and Ge precursors, thus changing the substitution ratio [Si]/([Si]+[Ge]) [37,118,119]. Based on these promising results, other isomorphic substitutions have been recently investigated, in particular by partially replacing Al$^{3+}$ by Fe$^{3+}$ in the outer wall [46,47,120,121]. Although these improvements do not imply significant change of the nanotube diameter (up to 1% of doping rate) [47], the effect of Fe incorporation acts as specific coordination centers for organic moieties and modifies the electronic properties of the nanotubes by decreasing the band gap, even for low substitution rates [45]. Other isomorphic substitutions have been proposed from the point of view of theory, either by replacing both Al and Si atoms by elements of groups III (Ga, In) and IV (C, Ge, Sn) [122] or by substituting the inner tetrahedrons by phosphorous and arsenite derivatives [123], but it remains to be implemented experimentally.

### 5.3.3. Surface Functionalization of Imogolite Nanotubes

An interesting feature of INTs is the high density of hydroxyl groups on their outer surface (~18 OH/nm$^2$), offering multiple binding sites for external surface functionalization. Depending on the intended application, it is often necessary to make INTs compatible with organic matrix. Hence, surface functionalization of INTs has been widely explored, as compiled in recent reviews [24,124,125]. An efficient way to modify a metal oxide surface is to use phosphonate derivatives [126]. This issue has been extensively studied [127–136] since 2001 and the first report of the chemisorption of octadecylphosphonic acid onto the outer surface of imogolite nanotubes [127]. Successful grafting of organic moieties can be evidenced by dispersing unmodified and modified nanotubes in organic solvents (Figure 5.8) [131,135,137].

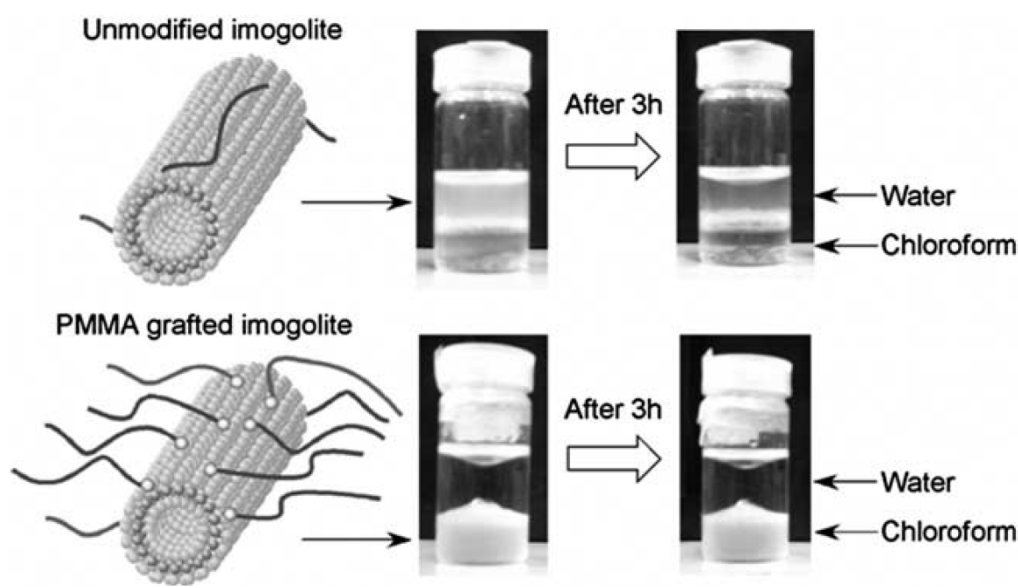

**Figure 5.8.** Illustration of the dispersion state of poly(methyl methacrylate) (PMMA) grafted aluminosilicate nanotubes and of unmodified ones in water/chloroform phase. Reprinted with permissions from [137].

Grafting of carboxylates [128,138,139] and sulfonates [140] groups have also been reported with similar results as with phosphonates. Otherwise, the use of silane as coupling agent appears to be controversial because the moieties remain hydrolytically labile and are totally removed after several days of dialysis [141,142]. Beyond chemisorption, chemical vapor deposition (CVD) [143,144] or γ-ray irradiation [145,146] have also been considered.

The inner hydrophilic cavity of INTs can also be functionalized by a post-synthesis chemical modification. Post-functionalization can be obtained with different silane agents such as (3-aminopropyl)triethoxysilane (APTES), methyltrimethoxysilane (MTMS) or trichlorosilane (TCIS) but also with acetyl chloride after dehydration of the hydrophilic INT samples at sufficiently high temperature [142,147]. However, the degree of inner surface substitution is lower than 35% [147]. On can also use direct, template-free, synthesis route by replacing, the initial alkoxide (prefiguring the tetrahedral layer) by a functionalized one (Figure 5.7c). Hydroxyl groups can be replaced by methyl [148] or amino [149] groups. Like for unmodified INTs, a dependence of the inner cavity diameter with the substitution of Si by Ge could be expected in hybrid nanotubes. Amara et al. [150] demonstrated that hybrid imogolite with methylated inner cavities can be obtained with Si to Ge endmembers (Figure 5.9) and that the inner diameter increases progressively with decreasing the ratio [Si]/([Si]+[Ge]), from 1.8 and 2.4 nm.

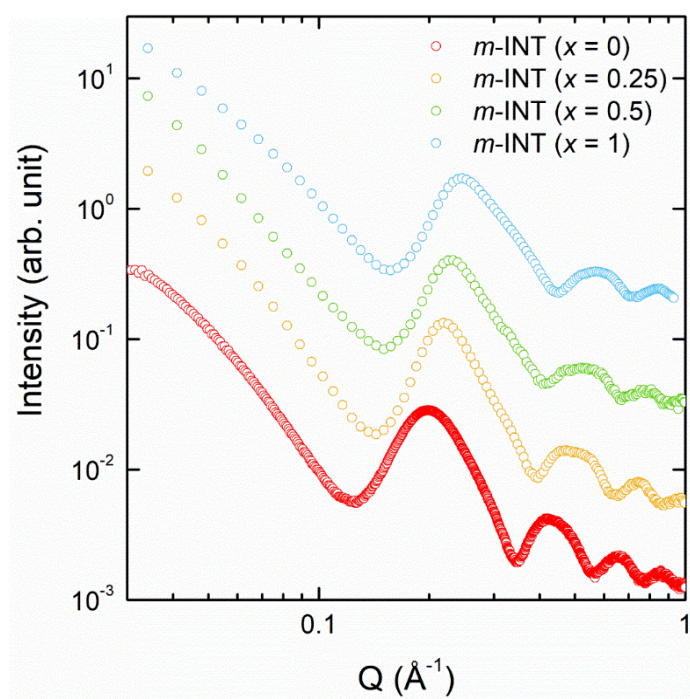

**Figure 5.9.** X-ray scattering (XRS) diagrams obtained on methyl-modified imogolite nanotubes of nominal composition $(OH)_3Al_2O_3Si_xGe_{1-x}(CH_3)$ prepared at different substitution ratio x = [Si]/([Si] + [Ge]). Analysis of XRS diagrams upon substitution of Si by Ge shows an increase of the nanotube diameter. Reprinted with permissions from [150].

Methylation of the inner cavity not only offers control over the surface properties but it also leads to a drastic structural change of the chirality of the nanotube (see section 5.2) [31]. Interestingly, Bahadori et al. recently considered the possibility to prepare novel Fe-doped imogolite nanotubes with methylated inner cavity, combining two type of isomorphic substitution [151].

In brief, the synthesis of imogolite-like nanostructures of the type $(OH)_3(Al_{2-x}M_x)O_3(Ge_{1-y}Si_y)R$ (with M = $Fe^{3+}$, $Ga^{3+}$…; R = –OH, –$CH_3$, –$CH_2NH_2$…) offers a unique system of 1D nanostructures with monodisperse sizes and adjustable aspect ratio.

## 5.4. Applications in Functional Materials
### 5.4.1. Reinforcement of Polymer Materials

With their well-defined morphologies and their easy surface functionalization, imogolite nanotubes have emerged during the last decades as serious candidates for reinforcing polymer nanocomposites by using various type of polymers in the form of films or membranes and very recently in the form of fibres (Table 5.1).

**Table 5.1.** Class and type of polymer used for imogolite-reinforced nanocomposites.

| Class of polymer | Type | Acronym | Final form | Ref. |
|---|---|---|---|---|
| Vinyl | Poly(vinyl alcohol) | PVOH | Film | [128,129,152–154] |
| | | | Membrane | [155–157] |
| | | | Fibre | [158] |
| | Poly(vinyl chloride) | PVC | Film | [159] |
| Acrylate | Poly(methyl methacrylate) | PMMA | Film | [137] |
| | Poly(acrylic acid) | PAA | Hydrogel | [145,146,160,161] |
| | Poly(hydroxyethyl acrylate) | PHEA | Hydrogel | [161] |
| Ester | Poly(ε-caprolactone) | PCL | Film | [162] |
| | | | Hydrogel | [163] |
| | Poly(lactic acid) | PLA | Hydrogel | [163] |
| | Poly(butylene succinate) | PBS | Hydrogel | [163] |
| | Polyamide | PA | Membrane | [164–166] |
| | | | Hydrogel | [161,167,168] |
| Others | Dicarboxilic acid | DA | Hydrogel | [138,139,169] |

Imogolite nanotubes have often been mixed with poly(vinyl alcohol) (PVOH) since PVOH/INT mixtures can be processed readily from aqueous solutions. PVOH/INT blend films were successfully prepared but the incorporation of imogolite nanotubes has only a slight influence on both Young modulus and tensile strength properties of the nanocomposite [152,153]. Instead of preparing films by a solution blend of imogolite powder, Yamamoto et al. made composites in situ by synthesis of INT in presence of the PVOH [129]. It improves the dispersion of INTs in the resulting film and enhances their optical and mechanical properties (Figure 5.10) [129].

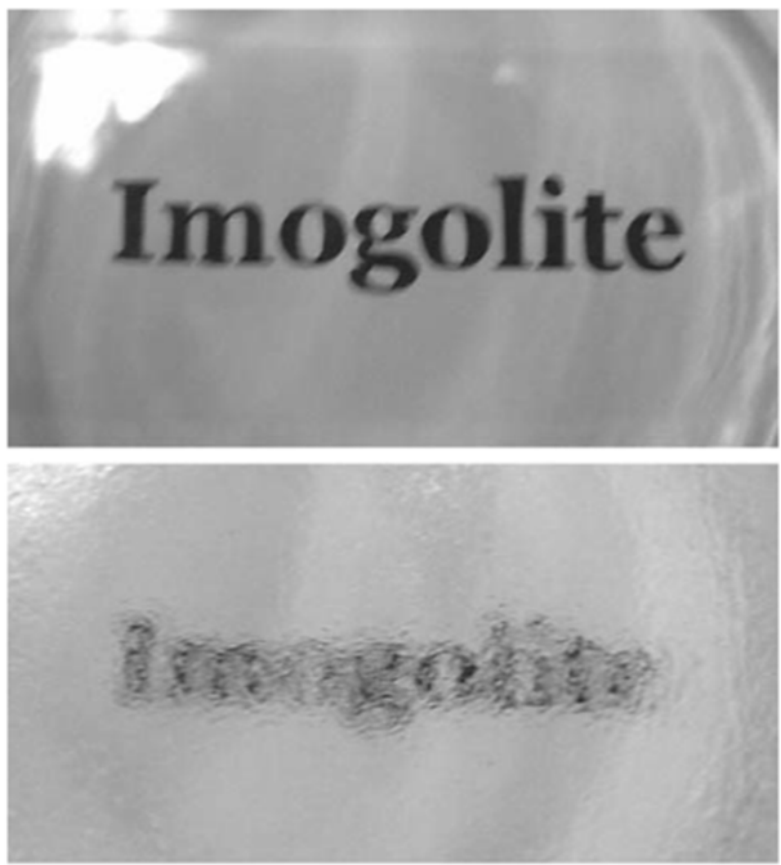

**Figure 5.10.** Optical transparency of in situ hybrid (upper) and blend (lower) imogolite/PVOH films. The thickness of films is ~ 100 mm. Reprinted with permissions from [129].

Similar enhancement of mechanical properties was also observed with poly(methyl methacrylate) (PMMA) [137] and poly(acrylic acid) (PAA) nanocomposites [145,146,160,161]. However, incorporation of PMMA grafted INT into poly(vinyl chloride) (PVC) films reveal that the reinforcement is strongly dependent on the interfacial adhesion between the functionalized nanotubes and the polymer matrix [159]. In addition to reinforcement in solid materials, the combination of imogolite with various polymers also yields homogeneous hydrogels exhibiting thixotropic behaviour, hierarchical ordering as well as reversible isotropic–anisotropic structural transitions in response to strain [138,139,161,163,167–169]. Imogolite hydrogels can also be prepared with biomacromolecules [170,171] for encapsulation and/or for sustained release of enzymes or model drugs [172,173].

### 5.4.2. Molecular Selectivity and Storage

Beyond their use as fillers for reinforced materials, imogolite nanotubes also present a promising potential for molecular confinement applications thanks to their well-defined diameters and interfaces. Imogolite nanotubes were proposed as an efficient molecular sieving material, especially for gas adsorption and storage [104,174–177]. Interestingly, the selective functionalization of the inner interface (see section 5.3.3) results in dramatically enhanced molecular selectivity over $CO_2/CH_4$ and $CO_2/N_2$ [149]. Methylated-modified imogolite nanotubes can also be used in aqueous suspensions for trapping organic molecules [135,150]. INTs could finally be employed as tuneable building blocks for filtration applications by reverse osmosis [178]. In combination with their compatibility with different polymer matrix, the fabrication of imogolite-based membranes has started to receive more intention recently. In this framework, several studies reported that addition of imogolite nanotubes in water filtration membranes (Figure 5.11) induces a substantial increase of the water flux while achieving significant ion exclusion [155,156,164–166]. However, the mechanisms involved in the enhancement of the membranes performance remain to be carefully investigated from the viewpoint of fundamental science. In particular, no studies has clearly established whether water molecules diffuse through or around imogolite nanotubes incorporated in the matrix.

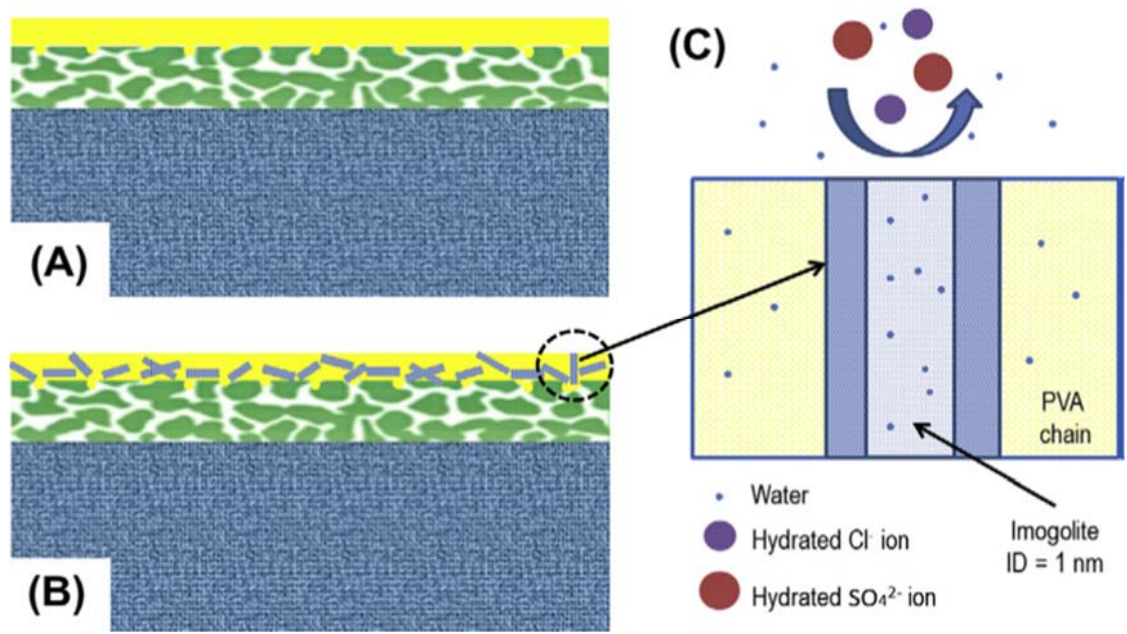

**Figure 5.11.** Schematic diagram of (a) pristine and (b) modified filtration membranes by incorporation of aluminosilicate imogolite nanotubes. (c) Mechanism proposed for water filtration and salt rejection (ID: internal diameter). Reprinted with permissions from [156].

### 5.4.3. Environment Field

Imogolite nanotubes were recognized as good adsorbents very early on, especially in environmental context where they present high capacity to bind anions and metallic cations. Parfitt et al. [72] reported that natural imogolite (pumice beds) are capable of fixing large amounts of phosphates up to 120-250 µmol.g$^{-1}$. They suggested that the adsorption mechanism is related to a ligand-exchange reaction at the surface of imogolite nanotubes. Similar mechanisms have been invoked for the adsorption of metallic cations ($Cu^{2+}$, $Pb^{2+}$, $UO_2^{2+}$) on the outer surface of INTs [84,179]. On the other hand, Levard et al. [180] revealed incorporation of Ni species into the octahedral vacancies of the external INT wall.

The immobilization of different noble metals nanoparticles on specific sites of the external surfaces was also investigates by Liz-Marzan and Philipse [91]. In that case, imogolite nanotubes in aqueous media act as a solid support, being effective to stabilize the formation of Pt, Au, Ag and bimetallic nanoparticles [181,182]. These results pave the way towards potential uses of imogolite nanotubes for water treatment [183–186], antimicrobial agents [187,188], transparent dressings [189] and for catalysis applications [190].

### 5.4.4. Catalysis

It was suggested early that INT may act as a catalyst [51]. But, despite the many progresses achieved in the synthesis of imogolite nanostructures, this field remains relatively unexplored. The first published example is probably the one on the coupling of Cu species with INT [190] that induces a higher activity in the decomposition of tert-butyl hydroperoxide than with unloaded imogolite or with Cu–$SiO_2$ materials. Katsumata et al. investigated complex $Cu^{2+}$-grafted $TiO_2$/INT composites which exhibit very efficient absorption and photodecomposition rate of acetaldehyde [191]. Imogolite nanotubes and their derivatives have also been used for hydroxylation of olefins [192], catalytic oxidation of aromatic hydrocarbon [193] isomerization of glucose to fructose [194] or photo-degradation of azo-dyes [151,186], although using dyes for evaluating photocatalytic properties remain controversial [195].

On the theoretical background, DFT calculations reveal a most interesting feature specific to imogolite structure. INTs bear a tunable permanent polarization of their wall and a real-space separation of the valence and conduction band edges is highlighted [38,39]. Characterization of the nanotube band separation and optical properties reveal the occurrence of near UV inside–outside charge-transfer excitations, which may be effective for enhanced photocatalytic activity [44,196]. In spite of previous use of INTs as catalyst support (see section 5.4.3) and preliminary results on the performance of INTs for photo-catalytic decomposition of organic dyes [151,186], to date this subject is still to be explored.

### 5.5. Environmental Health and Safety

Only few experimental studies have been performed on the cytotoxicity and genotoxicity of natural and synthetic imogolite nanotubes compared to carbon nanotubes for instance. In vitro tests performed on aluminosilicate nanotubes reveal that imogolite are internalised into murine macrophage cells and induce a low dose and time-dependent cytotoxicity, but to a lesser extent than single- and multi-walled carbon nanotubes [197]. A high toxicity effect of INT to lung cell cultures was reported by Farmer et al 1983 but without mentioning the dose administered [51]. Recent investigations do not show any genotoxic effect except for a decrease in DNA integrity observed in airway epithelial cells and for the highest dose (80 µg.cm$^{-2}$) [197]. By contrast, Liu et al. investigated the effect of Ge-imogolite nanotubes (average length: 6 and 51 nm long) on human fibroblast culture [198]. The authors observed higher internalization rates in the case of low aspect ratio Ge-INTs without decreasing the cell viability. However, these short nanotubes seem to induce significant lesions of DNA, mainly through oxidative stress while INTs with higher aspect ratio exhibit a weaker genotoxicity [198]. To go further, the respiratory biopersistence and lung toxicity of Ge-SWINT and Ge-DWINT with length of ~70 nm were evaluated in vivo and compared with crocidolite particles [199]. The results showed that Ge-INT are biopersistent and promote genotoxicity, sustained inflammation and fibrosis in lung cells, with effects more severe than those induced by crocidolite [199]. Care should be exercised in handling powders or freeze-dried imogolite samples, which are readily dispersed in air.

Attention should also be paid to the fact that the biological medium can strongly modify the physico-chemical properties of imogolite. Indeed, hydroxyl groups on the outer wall of imogolite nanotubes may interact either by bonding ionic species [73,79], present in the biological medium or by interacting directly with phosphonic acid groups of biological polymers [170]. Future research is necessary to confirm these hypotheses and they will rely on the recent progress made on the synthesis of various imogolite-like nanostructures with well-defined morphologies.

**5.6. Suggestions and Future Prospects**

We have shown in the previous sections that imogolite is naturally present but widely dispersed in environment, thus limiting its applications. The major breakthrough for using INTs was certainly their easy synthesis by sol-gel methods under mild conditions, a prerequisite for developing green materials. Such approach induces some drawbacks. Indeed, in the case of aluminosilicate imogolite nanotubes, the starting precursors are easily accessible and inexpensive, but often yielding low quantities of nanotubes. The situation is reversed in the case of aluminogermanate INTs.

It is now possible designing innovative INTs with modular interfaces in a predictive way, which has recently led to some progress the field of imogolite-based functional materials. Imogolite-like nanotubes can be considered as a flexible building block with multipurpose applications (nanocomposites, stimuli-responsive materials, (photo)catalysis, ion adsorption, molecular separation and storage…). It is interesting to note also that most of the applications explored up to now do not fully exploit all properties (controlled morphology, tuneable interfaces and colloidal proeperties) of these peculiar nanostructures. For instance, a future trend could be to combine the well-defined morphology of INTs with their ease to self-organize to produce novel nanocomposites and artificial ion channel devices, as was done recently with carbon nanotubes [200,201]. Finally, recent DFT calculations [44,196] showed that imogolite nanotubes display a real-space separation of the valence and conduction bands, which may be effective for electron-hole pair (e*-h) separation and enhanced photocatalytic activity, an interesting alternative to photoferroelectrics materials [202–204]. Coupled with molecular confinement, the promising use of polarized 1D nanotubes for photocatalysis applications deserves to be explored experimentally.

**Acknowledgement**

We would like to thank our collaborators who share our interest in imogolite nanotubes and in their properties: Geoffrey Monet (PhD student), Rajesh Bandhary (post-doctoral researcher), Stéphane Rols, Stéphan Rouzière, Pierre Levitz, Laurent Michot, Anne-Laure Rollet, Gilberto Teobaldi and Milo Shaffer.

**References**

[1] N. Yoshinaga, S. Aomine, Imogolite in some Ando soils, Soil Sci. Plant Nutr. 8 (1962) 22–29.

[2] N. Yoshinaga, S. Aomine, Allophane in some Ando soils, Soil Sci. Plant Nutr. 8 (1962) 6–13.

[3] M. Nanzyo, H. Kanno, Non-crystalline Inorganic Constituents of Soil, in: Inorganic Constituents in Soil, Springer, 2018: pp. 59–95.

[4] M. Fleischer, New Mineral Names, American Mineralogist: Journal of Earth and Planetary Materials 48 (1963) 433–436.

[5] S.W. Bailey, Summary of national and international recommendations on clay mineral nomenclature, Clays Clay Miner. 19 (1971) 129–132.

[6] N. Miyauchi, S. Aomine, Mineralogy of gel-like substance in the pumice bed in Kanuma and Kitakami districts, Soil Sci. Plant Nutr. 12 (1966) 19–22.

[7] von G. Jaritz, Ein Vorkommen von Imogolit in Bimsböden Westdeutschlands, Zeitschrift Für Pflanzenernährung Und Bodenkunde 117 (1967) 65–77.

[8] N. Yoshinaga, Identification of imogolite in the filmy gel materials in the imaichi and shichihonzakura purlice beds, Soil Sci. Plant Nutr. 14 (1968) 238–246.

[9] N. Yoshinaga, M. Yamaguchi, Occurrence of imogolite as gel film in the pumice and scoria deds of western and central Honshu and Hokkaido, Soil Sci. Plant Nutr. 16 (1970) 215–223.

[10] E. Besoain, Imogolite in volcanic soils of Chile, Geoderma 2 (1969) 151–169.

[11] C. Gense, Conditions de formation de l'imogolite dans les produits d'altération de roches volcaniques basiques de l'Ile de la Réunion, Bulletin Du Groupe Français Des Argiles 25 (1973) 79–84.

[12] N. Yoshinaga, J.M. Tait, R. Soong, Occurrence of imogolite in some volcanic ash soils of New Zealand, Clay Miner. 10 (1973) 127–130.

[13] R.L. Parfitt, Imogolite from New Guinea, Clays Clay Miner. 22 (1974) 369–371.

[14] K. Wada, Allophane and imogolite, in: Developments in Sedimentology, Elsevier, 1978: pp. 147–187.

[15] P. Violante, J. Tait, Identification of imogolite in some volcanic soils from Italy, Clay Miner. 14 (1979) 155–158.

[16] M. Panichini, R. Neculman, R. Godoy, N. Arancibia-Miranda, F. Matus, Understanding carbon storage in volcanic soils under selectively logged temperate rainforests, Geoderma 302 (2017) 76–88.

[17] J. Tait, N. Yoshinaga, B. Mitchell, Occurrence of imogolite in some scottish soils, Soil Sci. Plant Nutr. 24 (1978) 145–151.

[18] G. Ross, H. Kodama, Evidence for imogolite in canadian soils, Clays Clay Miner. 27 (1979) 297–300.

[19] R. Dahlgren, F. Ugolini, Formation and stability of imogolite in a tephritic spodosol, Cascade Range, Washington, USA, Geochim Cosmochim Acta 53 (1989) 1897–1904.

[20] J.P. Gustafsson, D.G. Lumsdon, M. Simonsson, Aluminium solubility characteristics of spodic B horizons containing imogolite-type materials, Clay Miner. 33 (1998) 77–86.

[21] J.P. Gustafsson, P. Bhattacharya, E. Karltun, Mineralogy of poorly crystalline aluminium phases in the B horizon of Podzols in southern Sweden, Appl. Geochem. 14 (1999) 707–718.

[22] J.L. Bishop, E.B. Rampe, Evidence for a changing Martian climate from the mineralogy at Mawrth Vallis, Earth Planet. Sci. Lett. 448 (2016) 42–48.

[23] J.L. Bishop, A.G. Fairén, J.R. Michalski, L. Gago-Duport, L.L. Baker, M.A. Velbel, C. Gross, E.B. Rampe, Surface clay formation during short-term warmer and wetter conditions on a largely cold ancient Mars, Nat. Astron. 2 (2018) 206.

[24] W. Ma, W.O. Yah, H. Otsuka, A. Takahara, Surface functionalization of aluminosilicate nanotubes with organic molecules, Beilstein J. Nanotech. 3 (2012) 82–100.

[25] N. Yoshinaga, H. t Yotsumoto, K. Ibe, An electron microscopic study of soil allophane with an ordered structure, American Mineralogist: Journal of Earth and Planetary Materials 53 (1968) 319–323.

[26] J.D. Russell, W.J. McHardy, A.R. Fraser, Imogolite: a unique aluminosilicate, Clay Miner. 8 (1969) 87–99.

[27] K. Wada, N. Yoshinaga, K. Yotsumoto, High resolution electron micrographs of imogolite, Clay Miner. 8 (1970) 487–489.

[28] K. Wada, A structural scheme of soil allophane, American Mineralogist: Journal of Earth and Planetary Materials 52 (1967) 690–708.

[29] K. Wada, N. Yoshinaga, The structure of "imogolite," American Mineralogist: Journal of Earth and Planetary Materials 54 (1969) 50–71.

[30] P.D.G. Cradwick, K. Wada, J. Russell, N. Yoshinaga, C. Masson, V. Farmer, Imogolite, a hydrated aluminum silicate of tubular structure, Nature-Physical Science. 240 (1972) 187–189.

[31] G. Monet, M.S. Amara, S. Rouzière, E. Paineau, Z. Chai, J.D. Elliott, E. Poli, L.-M. Liu, G. Teobaldi, P. Launois, Structural resolution of inorganic nanotubes with complex stoichiometry, Nat. Commun. 9 (2018) 2033.

[32] E. Hernandez, C. Goze, P. Bernier, A. Rubio, Elastic properties of C and B x C y N z composite nanotubes, Phys. Rev. Lett. 80 (1998) 4502.

[33] G. Seifert, H. Terrones, M. Terrones, G. Jungnickel, T. Frauenheim, Structure and electronic properties of MoS 2 nanotubes, Phys. Rev. Lett. 85 (2000) 146.

[34] K. Tamura, K. Kawamura, Molecular dynamics modeling of tubular aluminum silicate: Imogolite, J. Phys. Chem. B 106 (2002) 271–278.

[35] S. Konduri, S. Mukherjee, S. Nair, Strain energy minimum and vibrational properties of single-walled aluminosilicate nanotubes, Phys. Rev. B 74 (2006) 033401.

[36] L. Guimaraes, A.N. Enyashin, J. Frenzel, T. Heine, H.A. Duarte, G. Seifert, Imogolite nanotubes: Stability, electronic, and mechanical properties, ACS Nano 1 (2007) 362–368.

[37] S. Konduri, S. Mukherjee, S. Nair, Controlling nanotube dimensions: Correlation between composition, diameter, and internal energy of single-walled mixed oxide nanotubes, ACS Nano 1 (2007) 393–402.

[38] G. Teobaldi, N.S. Beglitis, A.J. Fisher, F. Zerbetto, A.A. Hofer, Hydroxyl vacancies in single-walled aluminosilicate and aluminogermanate nanotubes, J. Phys. Condens. Mat. 21 (2009) 195301.

[39] E. Poli, J.D. Elliott, N.D.M. Hine, A.A. Mostofi, G. Teobaldi, Large-scale density functional theory simulation of inorganic nanotubes: a case study on Imogolite nanotubes, Mater. Res. Innov. 19 (2015) S272–S282.

[40] E. Joussein, S. Petit, J. Churchman, B. Theng, D. Righi, B. Delvaux, Halloysite clay minerals–a review, Clay Miner. 40 (2005) 383–426.

[41] P. Yuan, D. Tan, F. Annabi-Bergaya, Properties and applications of halloysite nanotubes: recent research advances and future prospects, Appl. Clay Sci. 112 (2015) 75–93.

[42] L.A. Bursill, J.L. Peng, L.N. Bourgeois, Imogolite: an aluminosilicate nanotube material, Philosophical Magazine A-Physics of Condensed Matter Structure Defects and Mechanical Properties 80 (2000) 105–117.

[43] M.P. Lourenco, L. Guimaraes, M.C. da Silva, C. de Oliveira, T. Heine, H.A. Duarte, Nanotubes with well-defined structure: single- and double-walled imogolites, J. Phys. Chem. C 118 (2014) 5945–5953.

[44] J.D. Elliott, E. Poli, I. Scivetti, L.E. Ratcliff, L. Andrinopoulos, J. Dziedzic, N.D.M. Hine, A.A. Mostofi, C.-K. Skylaris, P.D. Haynes, G. Teobaldi, Chemically selective alternatives to photoferroelectrics for polarization-enhanced photocatalysis: the untapped potential of hybrid inorganic nanotubes, Adv. Sci. 4 (2017) 1600153.

[45] F. Alvarez-Ramirez, First principles studies of Fe-containing aluminosilicate and aluminogermanate nanotubes, J. Chem. Theory Comput. 5 (2009) 3224–3231.

[46] A. Avellan, C. Levard, N. Kumar, J. Rose, L. Olivi, A. Thill, P. Chaurand, D. Borschneck, A. Masion, Structural incorporation of iron into Ge-imogolite nanotubes: a promising step for innovative nanomaterials, RSC Adv. 4 (2014) 49827–49830.

[47] E. Shafia, S. Esposito, M. Manzoli, M. Chiesa, P. Tiberto, G. Barrera, G. Menard, P. Allia, F.S. Freyria, E. Garrone, B. Bonelli, Al/Fe isomorphic substitution versus $Fe_2O_3$ clusters formation in Fe-doped aluminosilicate nanotubes (imogolite), J. Nanopart. Res. 17 (2015) 336.

[48] S. Van der Gaast, K. Wada, S. Wada, Y. Kakuto, Small-angle X-ray powder diffraction, morphology, and structure of allophane and imogolite, Clays Clay Miner. 33 (1985) 237–243.

[49] K. MacKenzie, M. Bowden, I. Brown, R. Meinhold, Structure and thermal transformations of imogolite studied by $Si^{29}$ and $Al^{27}$ high-resolution solid-state nuclear magnetic-resonance, Clays Clay Miner. 37 (1989) 317–324.

[50] D.-Y. Kang, J. Zang, E.R. Wright, A.L. McCanna, C.W. Jones, S. Nair, Dehydration, dehydroxylation, and rehydroxylation of single-walled aluminosilicate nanotubes, ACS Nano 4 (2010) 4897–4907.
[51] V. Farmer, M. Adams, A. Fraser, F. Palmieri, Synthetic imogolite - properties, synthesis, and possible applications, Clay Miner. 18 (1983) 459–472.
[52] N. Donkai, T. Miyamoto, T. Kokubo, H. Tanei, Preparation of transparent mullite silica film by heat-treatment of imogolite, J. Mater. Sci. 27 (1992) 6193–6196.
[53] C. Zanzottera, A. Vicente, M. Armandi, C. Fernandez, E. Garrone, B. Bonelli, Thermal collapse of single-walled alumino-silicate nanotubes: transformation mechanisms and morphology of the resulting lamellar phases, J. Phys. Chem. C 116 (2012) 23577–23584.
[54] K.-H. Liou, N.-T. Tsou, D.-Y. Kang, Relationships among the structural topology, bond strength, and mechanical properties of single-walled aluminosilicate nanotubes, Nanoscale 7 (2015) 16222–16229.
[55] B. Lecouvet, J. Horion, C. D'haese, C. Bailly, B. Nysten, Elastic modulus of halloysite nanotubes, Nanotechnology 24 (2013) 105704.
[56] W. Li, G. Zhang, M. Guo, Y.-W. Zhang, Strain-tunable electronic and transport properties of $MoS_2$ nanotubes, Nano Res. 7 (2014) 518–527.
[57] S. Piperno, I. Kaplan-Ashiri, S.R. Cohen, R. Popovitz-Biro, H.D. Wagner, R. Tenne, E. Foresti, I.G. Lesci, N. Roveri, Characterization of geoinspired and synthetic chrysotile nanotubes by atomic force microscopy and transmission electron microscopy, Adv. Funct. Mater. 17 (2007) 3332–3338.
[58] G.I. Yucelen, R.P. Choudhury, J. Leisen, S. Nair, H.W. Beckham, Defect structures in aluminosilicate single-walled nanotubes: a solid-state nuclear magnetic resonance investigation, J. Phys. Chem. C 116 (2012) 17149–17157.
[59] M.C. da Silva, E.C. dos Santos, M.P. Lourenço, M.P. Gouvea, H.A. Duarte, Structural, electronic, and mechanical properties of inner surface modified imogolite nanotubes, Front. Mater.. 2 (2015) 16.
[60] K.-H. Liou, D.-Y. Kang, Defective Single-walled aluminosilicate nanotubes: structural stability and mechanical properties, ChemNanoMat 2 (2016) 189–195.
[61] B. Creton, D. Bougeard, K.S. Smirnov, J. Guilment, O. Poncelet, Molecular dynamics study of hydrated imogolite - 2. Structure and dynamics of confined water, Phys. Chem. Chem. Phys. 10 (2008) 4879–4888.
[62] M.S. Amara, S. Rouziere, E. Paineau, M. Bacia-Verloop, A. Thill, P. Launois, Hexagonalization of aluminogermanate imogolite nanotubes organized into closed-packed bundles, J. Phys. Chem. C 118 (2014) 9299–9306.
[63] A.P. Philipse, A.M. Wierenga, On the density and structure formation in gels and clusters of colloidal rods and fibers, Langmuir 14 (1998) 49–54.
[64] N. Donkai, H. Inagaki, K. Kajiwara, H. Urakawa, M. Schmidt, Dilute-solution properties of imogolite, Makromolekulare Chemie-Macromolecular Chemistry and Physics 186 (1985) 2623–2638.
[65] K. Kajiwara, N. Donkai, Y. Hiragi, H. Inagaki, Lyotropic mesophase of imogolite,1. Effect of polydispersity on phase-diagram, Makromolekulare Chemie-Macromolecular Chemistry and Physics 187 (1986) 2883–2893.

[66] K. Kajiwara, N. Donkai, Y. Fujiyoshi, H. Inagaki, Lyotropic mesophase of imogolite. 2. Microscopic observation of imogolite mesophase, Makromolekulare Chemie-Macromolecular Chemistry and Physics 187 (1986) 2895–2907.

[67] N. Donkai, H. Hoshino, K. Kajiwara, T. Miyamoto, Lyotropic mesophase of imogolite. 3. Observation of liquid-crystal structure by scanning electron and novel polarized optical microscopy, Makromolekulare Chemie-Macromolecular Chemistry and Physics 194 (1993) 559–580.

[68] P. Levitz, M. Zinsmeister, P. Davidson, D. Constantin, O. Poncelet, Intermittent Brownian dynamics over a rigid strand: Heavily tailed relocation statistics in a simple geometry, Phys. Rev. E 78 (2008) 030102.

[69] E. Paineau, M.-E.M. Krapf, M.-S. Amara, N.V. Matskova, I. Dozov, S. Rouziere, A. Thill, P. Launois, P. Davidson, A liquid-crystalline hexagonal columnar phase in highly-dilute suspensions of imogolite nanotubes, Nat. Commun. 7 (2016) 10271.

[70] L. Ramos, P. Fabre, Swelling of a lyotropic hexagonal phase by monitoring the radius of the cylinders, Langmuir 13 (1997) 682–686.

[71] E. Grelet, Hard-rod behavior in dense mesophases of semiflexible and rigid charged viruses, Phys. Rev. X 4 (2014) 021053.

[72] R. Parfitt, A. Thomas, R. Atkinson, R. Smart, Adsorption of phosphate on imogolite, Clays Clay Miner. 22 (1974) 455–456.

[73] J. Harsh, S. Traina, J. Boyle, Y. Yang, Adsorption of cations on imogolite and their effect on surface-charge characteristics, Clays Clay Miner. 40 (1992) 700–706.

[74] C. Su, J. Harsh, P. Bertsch, Sodium and chloride sorption by imogolite and allophanes, Clays Clay Miner. 40 (1992) 280–286.

[75] R.L. Parfitt, Allophane and imogolite: role in soil biogeochemical processes, Clay Miner. 44 (2009) 135–155.

[76] J. Karube, K. Nakaishi, H. Sugimoto, M. Fujihira, Electrophoretic behavior of imogolite under alkaline conditions, Clays Clay Miner. 40 (1992) 625–628.

[77] C. Su, J. Harsh, The electrophoretic mobility of imogolite and allophane in the presence of inorganic anions and citrate, Clays Clay Miner. 41 (1993) 461–471.

[78] H. Tsuchida, S. Ooi, K. Nakaishi, Y. Adachi, Effects of pH and ionic strength on electrokinetic properties of imogolite, Coll. Surf. A 265 (2005) 131–134.

[79] E. Paineau, M.S. Amara, G. Monet, V. Peyre, S. Rouzière, P. Launois, Effect of ionic strength on the bundling of metal oxide imogolite nanotubes, J. Phys. Chem. C 121 (2017) 21740–21749.

[80] J.P. Gustafsson, The surface chemistry of imogolite, Clays Clay Miner. 49 (2001) 73–80.

[81] P. Pasbakhsh, G.J. Churchman, J.L. Keeling, Characterisation of properties of various halloysites relevant to their use as nanotubes and microfibre fillers, Appl. Clay Sci. 74 (2013) 47–57.

[82] P. Pasbakhsh, G.J. Churchman, Natural mineral nanotubes: properties and applications, CRC Press, 2015.

[83] V. Farmer, A. Fraser, J. Tait, Synthesis of imogolite - tubular aluminum silicate polymer, J. Chem. Soc. Chem. Commun. (1977) 462–463.

[84] L. Denaix, I. Lamy, J.Y. Bottero, Structure and affinity towards $Cd^{2+}$, $Cu^{2+}$, $Pb^{2+}$ of synthetic colloidal amorphous aluminosilicates and their precursors, Coll. Surf. A 158 (1999) 315–325.

[85] S. Mukherjee, K. Kim, S. Nair, Short, highly ordered, single-walled mixed-oxide nanotubes assemble from amorphous nanoparticles, J. Am. Chem. Soc. 129 (2007) 6820–6826.

[86] C. Levard, J. Rose, A. Thill, A. Masion, E. Doelsch, P. Maillet, O. Spalla, L. Olivi, A. Cognigni, F. Ziarelli, J.-Y. Bottero, Formation and growth mechanisms of imogolite-like aluminogermanate nanotubes, Chem. Mater. 22 (2010) 2466–2473.

[87] G.I. Yucelen, R.P. Choudhury, A. Vyalikh, U. Scheler, H.W. Beckham, S. Nair, Formation of single-walled aluminosilicate nanotubes from molecular precursors and curved nanoscale intermediates, J. Am. Chem. Soc. 133 (2011) 5397–5412.

[88] G.I. Yucelen, D.-Y. Kang, R.C. Guerrero-Ferreira, E.R. Wright, H.W. Beckham, S. Nair, Shaping single-walled metal oxide nanotubes from precursors of controlled curvature, Nano Lett. 12 (2012) 827–832.

[89] A. Thill, P. Picot, L. Belloni, A mechanism for the sphere/tube shape transition of nanoparticles with an imogolite local structure (imogolite and allophane), Appl. Clay Sci. 141 (2017) 308–315.

[90] S. Barrett, P. Budd, C. Price, The synthesis and characterization of imogolite, Eur. Polym. J. 27 (1991) 609–612.

[91] L.L. Marzan, A.P. Philipse, Synthesis of platinum nanoparticles in aqueous host dispersions of inorganic (imogolite) rods, Coll. Surf. A 90 (1994) 95–109.

[92] Z. Abidin, N. Matsue, T. Henmi, A new method for nano tube imogolite synthesis, Jpn J. Appl. Phys. 47 (2008) 5079–5082.

[93] C. Levard, A. Masion, J. Rose, E. Doelsch, D. Borschneck, C. Dominici, F. Ziarelli, J.-Y. Bottero, Synthesis of imogolite fibers from decimolar concentration at low temperature and ambient pressure: a promising route for inexpensive nanotubes, J. Am. Chem. Soc. 131 (2009) 17080–17081.

[94] A. Chemmi, J. Brendle, C. Marichal, B. Lebeau, A novel fluoride route for the synthesis of aluminosilicate nanotubes, Nanomaterials 3 (2013) 117–125.

[95] A. Chemmi, J. Brendle, C. Marichal, B. Lebeau, Key steps influencing the formation of aluminosilicate nanotubes by the fluoride route, Clays Clay Miner. 63 (2015) 132–143.

[96] N. Arancibia-Miranda, M. Escudey, R. Ramirez, R.I. Gonzalez, A.C.T. van Duin, M. Kiwi, Advancements in the synthesis of building block materials: experimental evidence and modeled interpretations of the effect of Na and K on imogolite synthesis, J. Phys. Chem. C 121 (2017) 12658–12668.

[97] H. Yang, C. Wang, Z. Su, Growth mechanism. of synthetic imogolite nanotubes, Chem. Mater. 20 (2008) 4484–4488.

[98] G.I. Yucelen, D.-Y. Kang, I. Schmidt-Krey, H.W. Beckham, S. Nair, A generalized kinetic model for the formation and growth of single-walled metal oxide nanotubes, Chem. Eng. Sci. 90 (2013) 200–212.

[99] S. Wada, Imogolite synthesis at 25-degrees, Clays Clay Miner. 35 (1987) 379–384.

[100] H. Lee, Y. Jeon, Y. Lee, S.U. Lee, A. Takahara, D. Sohn, Thermodynamic control of diameter-modulated aluminosilicate nanotubes, J. Phys. Chem. C 118 (2014) 8148–8152.

[101] C.H. Lam, A.-C. Yang, H.-Y. Chi, K.-Y. Chan, C.-C. Hsieh, D.-Y. Kang, Microwave-assisted synthesis of highly monodispersed single-walled aluminosilicate nanotubes, ChemistrySelect 1 (2016) 6212–6216.

[102] A. Avellan, C. Levard, C. Chaneac, D. Borschneck, F.R.A. Onofri, J. Rose, A. Masion, Accelerated microwave assisted synthesis of alumino-germanate imogolite nanotubes, RSC Adv. 6 (2016) 108146–108150.

[103] V.C. Farmer, A.R. Fraser, Synthetic imogolite, a tubular hydroxyaluminium silicate, in: Developments in Sedimentology, Elsevier, 1979: pp. 547–553.

[104] F. Ohashi, S. Tomura, K. Akaku, S. Hayashi, S.I. Wada, Characterization of synthetic imogolite nanotubes as gas storage, J. Mater. Sci. 39 (2004) 1799–1801.

[105] T. Hongo, J. Sugiyama, A. Yamazaki, A. Yamasaki, Synthesis of imogolite from rice husk ash and evaluation of its acetaldehyde adsorption ability, Ind. Eng. Chem. Res. 52 (2013) 2111–2115.

[106] C. Levard, A. Masion, J. Rose, E. Doelsch, D. Borschneck, L. Olivi, P. Chaurand, C. Dominici, F. Ziarelli, A. Thill, P. Maillet, J.Y. Bottero, Synthesis of Ge-imogolite: influence of the hydrolysis ratio on the structure of the nanotubes, Phys. Chem. Chem. Phys. 13 (2011) 14516–14522.

[107] S. Mukherjee, V.A. Bartlow, S. Nair, Phenomenology of the growth of single-walled aluminosilicate and aluminogermanate nanotubes of precise dimensions, Chem. Mater. 17 (2005) 4900–4909.

[108] P. Picot, Y.Y. Liao, E. Barruet, F. Gobeaux, T. Coradin, A. Thill, Exploring hybrid imogolite nanotubes formation via Si/Al stoichiometry control, Langmuir 34 (2018) 13225-13234.

[109] K. Inoue, P. Huang, Influence of citric-acid on the natural formation of imogolite, Nature 308 (1984) 58–60.

[110] K. Inoue, P.M. Huano, Influence of citric acid on the formation of short-range ordered aluminosilicates, Clays Clay Miner 33 (1985) 312-322.

[111] R. Nakanishi, S.-I. Wada, M. Suzuki, M. Maeda, Heat-induced gelation of hydroxy-aluminosilicate synthesized by instantaneous mixing of sodium silicate and aluminum chloride, J. Fac. Agr. Kyushu U. 52 (2007) 147–151.

[112] S. Wada, K. Wada, Effects of substitution of germanium for silicon in imogolite, Clays Clay Miner. 30 (1982) 123–128.

[113] C. Levard, J. Rose, A. Masion, E. Doelsch, D. Borschneck, L. Olivi, C. Dominici, O. Grauby, J.C. Woicik, J.-Y. Bottero, Synthesis of large quantities of single-walled aluminogermanate nanotube, J. Am. Chem. Soc. 130 (2008) 5862–5863.

[114] P. Maillet, C. Levard, E. Larquet, C. Mariet, O. Spalla, N. Menguy, A. Masion, E. Doelsch, J. Rose, A. Thill, Evidence of double-walled Al-Ge imogolite-like nanotubes. a cryo-TEM and SAXS investigation, J. Am. Chem. Soc. 132 (2010) 1208–1209.

[115] P. Maillet, C. Levard, O. Spalla, A. Masion, J. Rose, A. Thill, Growth kinetic of single and double-walled aluminogermanate imogolite-like nanotubes: an experimental and modeling approach, Phys. Chem. Chem. Phys. 13 (2011) 2682–2689.

[116] A. Thill, P. Maillet, B. Guiose, O. Spalla, L. Belloni, P. Chaurand, M. Auffan, L. Olivi, J. Rose, Physico-chemical control over the single- or double-wall structure of aluminogermanate imogolite-like nanotubes, J. Am. Chem. Soc. 134 (2012) 3780–3786.

[117] M.-S. Amara, E. Paineau, M. Bacia-Verloop, M.-E.M. Krapf, P. Davidson, L. Belloni, C. Levard, J. Rose, P. Launois, A. Thill, Single-step formation of micron long $(OH)_3Al_2O_3Ge(OH)$ imogolite-like nanotubes, Chem. Commun. 49 (2013) 11284–11286.

[118] A. Thill, B. Guiose, M. Bacia-Verloop, V. Geertsen, L. Belloni, How the diameter and structure of $(OH)_3Al_2O_3Si_xGe_{1-x}OH$ imogolite nanotubes are controlled by an adhesion versus curvature competition, J. Phys. Chem. C. 116 (2012) 26841–26849.

[119] F. Alvarez-Ramirez, Ab initio simulation of the structural and electronic properties of aluminosilicate and aluminogermanate natotubes with imogolite-like structure, Phys. Rev. B 76 (2007) 125421.

[120] M. Ookawa, Y. Inoue, M. Watanabe, M. Suzuki, T. Yamaguchi, Synthesis and characterization of Fe containing imogolite, Clay Sci. 12 (2006) 280–284.

[121] E. Shafia, S. Esposito, M. Armandi, M. Manzoli, E. Garrone, B. Bonelli, Isomorphic substitution of aluminium by iron into single-walled alumino-silicate nanotubes: A physico-chemical insight into the structural and adsorption properties of Fe-doped imogolite, Micropor. Mesopor. Mater. 224 (2016) 229–238.

[122] F. Alvarez-Ramirez, Theoretical study of $(OH)_3N_2O_3MOH$, M = C, Si, Ge, Sn and N = Al, Ga, In, with imogolite-like structure, J. Computat. Theor. Nanos. 6 (2009) 1120–1124.

[123] L. Guimaraes, Y.N. Pinto, M.P. Lourenco, H.A. Duarte, Imogolite-like nanotubes: structure, stability, electronic and mechanical properties of the phosphorous and arsenic derivatives, Phys. Chem. Chem. Phys. 15 (2013) 4303–4309.

[124] K. Shikinaka, Design of stimuli-responsive materials consisting of the rigid cylindrical inorganic polymer "imogolite", Polym. J. 48 (2016) 689–696.

[125] E. Paineau, Imogolite nanotubes: a flexible nanoplatform with multipurpose applications, Appl. Sci. 8 (2018) 1921.

[126] M.-A. Neouze, U. Schubert, Surface modification and functionalization of metal and metal oxide nanoparticles by organic ligands, Monatshefte Für Chemie-Chemical Monthly 139 (2008) 183–195.

[127] K. Yamamoto, H. Otsuka, S. Wada, A. Takahara, Surface modification of aluminosilicate nanofiber "imogolite", Chem. Lett. (2001) 1162–1163.

[128] K. Yamamoto, H. Otsuka, A. Takahara, S.I. Wada, Preparation of a novel (polymer/inorganic nanofiber) composite through surface modification of natural aluminosilicate nanofiber, J. Adhesion 78 (2002) 591–602.

[129] K. Yamamoto, H. Otsuka, S.I. Wada, D. Sohn, A. Takahara, Transparent polymer nanohybrid prepared by in situ synthesis of aluminosilicate nanofibers in poly(vinyl alcohol) solution, Soft Matter 1 (2005) 372–377.

[130] S. Park, Y. Lee, B. Kim, J. Lee, Y. Jeong, J. Noh, A. Takahara, D. Sohn, Two-dimensional alignment of imogolite on a solid surface, Chem. Commun. (2007) 2917–2919.

[131] B.H. Bac, Y. Song, M.H. Kim, Y.-B. Lee, I.M. Kang, Surface-modified aluminogermanate nanotube by OPA: Synthesis and characterization, Inorganic Chem. Commun. 12 (2009) 1045–1048.

[132] W. Ma, J. Kim, H. Otsuka, A. Takahara, Surface modification of individual imogolite nanotubes with alkyl phosphate from an aqueous solution, Chem. Lett. 40 (2011) 159–161.

[133] W.O. Yah, A. Irie, N. Jiravanichanun, H. Otsuka, A. Takahara, Molecular aggregation state and electrical properties of terthiophenes/imogolite nanohybrids, Bull. Chem. Soc. Jpn. 84 (2011) 893–902.

[134] W. Ma, H. Otsuka, A. Takahara, Poly(methyl methacrylate) grafted imogolite nanotubes prepared through surface-initiated ARGET ATRP, Chem. Commun. 47 (2011) 5813–5815.

[135] P. Picot, O. Tache, F. Malloggi, T. Coradin, A. Thill, Behaviour of hybrid inside/out Janus nanotubes at an oil/water interface. A route to self-assembled nanofluidics?, Faraday Discuss. 191 (2016) 391–406.

[136] M. Li, J.A. Brant, Dispersing surface-modified imogolite nanotubes in polar and non-polar solvents, J. Nanopart. Res. 20 (2018) 19.

[137] K. Yamamoto, H. Otsuka, S.I. Wada, D. Sohn, A. Takahara, Preparation and properties of [poly(methyl methacrylate)/imogolite] hybrid via surface modification using phosphoric acid ester, Polymer 46 (2005) 12386–12392.

[138] K. Shikinaka, K. Kaneda, S. Mori, T. Maki, H. Masunaga, Y. Osada, K. Shigehara, Direct evidence for structural transition promoting shear thinning in cylindrical colloid assemblies, Small 10 (2014) 1813–1820.

[139] K. Shikinaka, H. Kikuchi, T. Maki, K. Shigehara, H. Masunaga, H. Sato, Chiral-linkage-induced hierarchical ordering of colloidal achiral nanotubes in their thixotropic gel, Langmuir 32 (2016) 3665–3669.

[140] N. Jiravanichanun, K. Yamamoto, A. Irie, H. Otsuka, A. Takahara, Preparation of hybrid films of aluminosilicate nanofiber and conjugated polymer, Synthetic Met. 159 (2009) 885–888.

[141] L. Johnson, T. Pinnavaia, Silylation of a tubular aluminosilicate polymer (imogolite) by reaction with hydrolyzed ($\gamma$-aminopropyl)triethoxysilane, Langmuir 6 (1990) 307–311.

[142] L. Johnson, T. Pinnavaia, Hydrolysis of (γ-aminopropyl)triethoxysilane-silylated imogolite and formation of a silylated tubular silicate-layered nanocomposite, Langmuir 7 (1991) 2636–2641.
[143] Y. Lee, B. Kim, W. Yi, A. Takahara, D. Sohn, Conducting properties of polypyrrole coated imogolite, Bull. Korean Chem. Soc. 27 (2006) 1815–1818.
[144] S. Chang, J. Park, J. Jang, J. Lee, J. Lee, W. Yi, Effect of UV irradiation during synthesis of polypyrrole by a one-step deposition/polymerization process, J. Vac. Sci. Technol B 25 (2007) 670–673.
[145] H. Lee, J. Ryu, D. Kim, Y. Joo, S.U. Lee, D. Sohn, Preparation of an imogolite/poly(acrylic acid) hybrid gel, J. Colloid Interface Sci. 406 (2013) 165–171.
[146] J. Ryu, H. Kim, J. Kim, J. Ko, D. Sohn, Dynamic behavior of hybrid poly (acrylic acid) gel prepared by γ-ray irradiated imogolite, Coll. Surf. A 535 (2017) 166–174.
[147] D.-Y. Kang, J. Zang, C.W. Jones, S. Nair, Single-walled aluminosilicate nanotubes with organic-modified interiors, J. Phys. Chem. C 115 (2011) 7676–7685.
[148] I. Bottero, B. Bonelli, S.E. Ashbrook, P.A. Wright, W. Zhou, M. Tagliabue, M. Armandi, E. Garrone, Synthesis and characterization of hybrid organic/inorganic nanotubes of the imogolite type and their behaviour towards methane adsorption, Phys. Chem. Chem. Phys. 13 (2011) 744–750.
[149] D.-Y. Kang, N.A. Brunelli, G.I. Yucelen, A. Venkatasubramanian, J. Zang, J. Leisen, P.J. Hesketh, C.W. Jones, S. Nair, Direct synthesis of single-walled aminoaluminosilicate nanotubes with enhanced molecular adsorption selectivity, Nat. Commun. 5 (2014) 3342.
[150] M.S. Amara, E. Paineau, S. Rouziere, B. Guiose, M.-E.M. Krapf, O. Tache, P. Launois, A. Thill, Hybrid, tunable-diameter, metal oxide nanotubes for trapping of organic molecules, Chem. Mater. 27 (2015) 1488–1494.
[151] E. Bahadori, V. Vaiano, S. Esposito, M. Armandi, D. Sannino, B. Bonelli, Photo-activated degradation of tartrazine by $H_2O_2$ as catalyzed by both bare and Fe-doped methyl-imogolite nanotubes, Catal. Today 304 (2018) 199–207.
[152] H. Hoshino, T. Ito, N. Donkai, H. Urakawa, K. Kajiwara, Lyotropic mesophase formation in PVA/imogolite mixture, Polym. Bull. 29 (1992) 453–460.
[153] J.H. Choi, Y.W. Cho, W.S. Ha, W.S. Lyoo, C.J. Lee, B.C. Ji, S.S. Han, W.S. Yoon, Preparation and characterization of syndiotacticity-rich ultra-high molecular weight poly(vinyl alcohol) imogolite blend film, Polym. Int. 47 (1998) 237–242.
[154] A.-C. Yang, Y.-S. Li, C.H. Lam, H.-Y. Chi, I.-C. Cheng, D.-Y. Kang, Solution-processed ultra-low-k thin films comprising single-walled aluminosilicate nanotubes, Nanoscale 8 (2016) 17427–17432.
[155] D.-Y. Kang, H.M. Tong, J. Zang, R.P. Choudhury, D.S. Sholl, H.W. Beckham, C.W. Jones, S. Nair, Single-walled aluminosilicate nanotube/poly(vinyl alcohol) nanocomposite membranes, ACS Appl. Mater. Inter. 4 (2012) 965–976.
[156] G.N.B. Barona, M. Choi, B. Jung, High permeate flux of PVA/PSf thin film composite nanofiltration membrane with aluminosilicate single-walled nanotubes, J. Colloid Interface Sci. 386 (2012) 189–197.

[157] D.-Y. Kang, M.E. Lydon, G.I. Yucelen, C.W. Jones, S. Nair, Solution-processed ultrathin aluminosilicate nanotube-poly(vinyl alcohol) composite membranes with partial alignment of nanotubes, ChemNanoMat. 1 (2015) 102–108.

[158] C.-Y. Su, A.-C. Yang, J.-S. Jiang, Z.-H. Yang, Y.-S. Huang, D.-Y. Kang, C.-C. Hua, Properties of single-walled aluminosilicate nanotube/poly (vinyl alcohol) aqueous dispersions, J. Phys. Chem. B 122 (2018) 380–391.

[159] W. Ma, H. Otsuka, A. Takahara, Preparation and properties of PVC/PMMA-g-imogolite nanohybrid via surface-initiated radical polymerization, Polymer 52 (2011) 5543–5550.

[160] J. Ryu, J. Ko, H. Lee, T.-G. Shin, D. Sohn, Structural response of imogolite-poly(acrylic acid) hydrogel under deformation, Macromolecules 49 (2016) 1873–1881.

[161] K. Shikinaka, T. Yokoi, Y. Koizumi-Fujii, M. Shimotsuya, K. Shigehara, Robust imogolite hydrogels with tunable physical properties, RSC Adv. 5 (2015) 46493–46500.

[162] S. Miura, N. Teramoto, M. Shibata, Nanocomposites composed of poly($\varepsilon$-caprolactone) and oligocaprolactone-modified imogolite utilizing biomimetic chelating method, J. Polym. Res. 23 (2016) 19.

[163] K. Shikinaka, A. Abe, K. Shigehara, Nanohybrid film consisted of hydrophobized imogolite and various aliphatic polyesters, Polymer 68 (2015) 279–283.

[164] G.N.B. Barona, J. Lim, M. Choi, B. Jung, Interfacial polymerization of polyamide-aluminosilicate SWNT nanocomposite membranes for reverse osmosis, Desalination 325 (2013) 138–147.

[165] Y.-H. Pan, Q.-Y. Zhao, L. Gu, Q.-Y. Wu, Thin film nanocomposite membranes based on imogolite nanotubes blended substrates for forward osmosis desalination, Desalination 421 (2017) 160–168.

[166] M. Li, J.A. Brant, Synthesis of Polyamide Thin-film nanocomposite membranes using surface modified imogolite nanotubes, J. Membrane Sci. 563 (2018) 664-675.

[167] K. Shikinaka, Y. Koizumi, Y. Osada, K. Shigehara, Reinforcement of hydrogel by addition of fiber-like nanofiller, Polym. Adv. Techno. 22 (2011) 1212–1215.

[168] K. Shikinaka, Y. Koizumi, K. Kaneda, Y. Osada, H. Masunaga, K. Shigehara, Strain-induced reversible isotropic-anisotropic structural transition of imogolite hydrogels, Polymer 54 (2013) 2489–2492.

[169] K. Shikinaka, N. Taki, K. Kaneda, Y. Tominaga, Quasi-solid electrolyte: a thixotropic gel of imogolite and an ionic liquid, Chem. Commun. 53 (2017) 613–616.

[170] N. Jiravanichanun, K. Yamamoto, K. Kato, J. Kim, S. Horiuchi, W.-O. Yah, H. Otsuka, A. Takahara, Preparation and characterization of imogolite/DNA hybrid hydrogels, Biomacromolecules 13 (2012) 276–281.

[171] B. Thomas, T. Coradin, G. Laurent, R. Valentin, Z. Mouloungui, F. Babonneau, N. Baccile, Biosurfactant-mediated one-step synthesis of hydrophobic functional imogolite nanotubes, RSC Adv. 2 (2012) 426–435.

[172] K. Kato, K. Inukai, K. Fujikura, T. Kasuga, Effective encapsulation of laccase in an aluminium silicate nanotube hydrogel, New Journal of Chemistry. 38 (2014) 3591–3599. doi:10.1039/c4nj00080c.

[173] R. Gelli, S. Del Buffa, P. Tempesti, M. Bonini, F. Ridi, P. Baglioni, Enhanced formation of hydroxyapatites in gelatin/imogolite macroporous hydrogels, J. Colloid Interface Sci. 511 (2018) 145–154.

[174] J. Karube, Y. Abe, Water retention by colloidal allophane and imogolite with different charges, Clays Clay Miner. 46 (1998) 322–329.

[175] M.A. Wilson, G.S.H. Lee, R.C. Taylor, Benzene displacement on imogolite, Clays Clay Miner. 50 (2002) 348–351.

[176] B. Bonelli, I. Bottero, N. Ballarini, S. Passeri, F. Cavani, E. Garrone, IR spectroscopic and catalytic characterization of the acidity of imogolite-based systems, J. Catal. 264 (2009) 15–30.

[177] B. Bonelli, C. Zanzottera, M. Armandi, S. Esposito, E. Garrone, IR spectroscopic study of the acidic properties of alumino-silicate single-walled nanotubes of the imogolite type, Catal. Today. 218 (2013) 3–9.

[178] K.-H. Liou, D.-Y. Kang, L.-C. Lin, Investigating the potential of single-walled aluminosilicate nanotubes in water desalination, ChemPhysChem. 18 (2017) 179–183.

[179] Y. Arai, M. McBeath, J.R. Bargar, J. Joye, J.A. Davis, Uranyl adsorption and surface speciation at the imogolite-water interface: Self-consistent spectroscopic and surface complexation models, Geochim. Cosmochim. Acta 70 (2006) 2492–2509.

[180] C. Levard, E. Doelsch, J. Rose, A. Masion, I. Basile-Doelsch, O. Proux, J.-L. Hazemann, D. Borschneck, J.-Y. Bottero, Role of natural nanoparticles on the speciation of Ni in andosols of la Reunion, Geochim. Cosmochim. Acta 73 (2009) 4750–4760.

[181] L. Liz-Marzán, A. Philipse, Stable hydrosols of metallic and bimetallic nanoparticles immobilized on imogolite fibers, J. Phys. Chem. 99 (1995) 15120–15128.

[182] Y. Kuroda, K. Fukumoto, K. Kuroda, Uniform and high dispersion of gold nanoparticles on imogolite nanotubes and assembly into morphologically controlled materials, Appl. Clay Sci. 55 (2012) 10–17.

[183] J. Guilment, D. Martin, O. Poncelet, Hybrid organic-inorganic materials designed to clean wash water in photographic processing: Genesis of a sol-gel industrial product the Kodak Water Saving Treatment System, in: C. Sanchez, R.M. Laine, S. Yang, C.J. Brinker (Eds.), Organic/Inorganic Hybrid Materials-2002, 2002: pp. 217–222.

[184] D.L. Guerra, A.C. Batista, R.R. Viana, C. Airoldi, Adsorption of rubidium on raw and MTZ- and MBI-imogolite hybrid surfaces: An evidence of the chelate effect, Desalination 275 (2011) 107–117.

[185] N. Arancibia-Miranda, M. Escudey, C. Pizarro, J.C. Denardin, M. Teresa Garcia-Gonzalez, J.D. Fabris, L. Charlet, Preparation and characterization of a single-

walled aluminosilicate nanotube-iron oxide composite: Its applications to removal of aqueous arsenate, Mater. Res. Bull. 51 (2014) 145–152.

[186] E. Shafia, S. Esposito, M. Armandi, E. Bahadori, E. Garrone, B. Bonelli, Reactivity of bare and Fe-doped alumino-silicate nanotubes (imogolite) with $H_2O_2$ and the azo-dye Acid Orange 7, Catal. Today. 277 (2016) 89–96.

[187] D.A. Geraldo, N. Arancibia-Miranda, N.A. Villagra, G.C. Mora, R. Arratia-Perez, Synthesis of CdTe QDs/single-walled aluminosilicate nanotubes hybrid compound and their antimicrobial activity on bacteria, J. Nanopart. Res. 14 (2012) 1286.

[188] G.I. Yucelen, R.E. Connell, J.R. Terbush, D.J. Westenberg, F. Dogan, Synthesis and immobilization of silver nanoparticles on aluminosilicate nanotubes and their antibacterial properties, Appl. Nanosci. 6 (2016) 607–614.

[189] Y.J. Lerat, O.J. Poncelet, Dressing and antiseptic agent containing silver, (2008) US Patent, 7,323,614

[190] S. Imamura, T. Kokubu, T. Yamashita, Y. Okamoto, K. Kajiwara, H. Kanai, Shape-selective copper-loaded imogolite catalyst, J. Catal. 160 (1996) 137–139.

[191] K. Katsumata, X. Hou, M. Sakai, A. Nakajima, A. Fujishima, N. Matsushita, K.J.D. MacKenzie, K. Okada, Visible-light-driven photodegradation of acetaldehyde gas catalyzed by aluminosilicate nanotubes and Cu(II)-grafted $TiO_2$ composites, Appl. Catal. B-Environ. 138 (2013) 243–252.

[192] X. Qi, H. Yoon, S.-H. Lee, J. Yoon, S.-J. Kim, Surface-modified imogolite by 3-APS-$OsO_4$ complex: Synthesis, characterization and its application in the dihydroxylation of olefins, J. Ind. Eng. Chem. 14 (2008) 136–141.

[193] M. Ookawa, Y. Takata, M. Suzuki, K. Inukai, T. Maekawa, T. Yamaguchi, Oxidation of aromatic hydrocarbons with $H_2O_2$ catalyzed by a nano-scale tubular aluminosilicate, Fe-containing imogolite, Res. Chem. Intermediates. 34 (2008) 679–685.

[194] N. Olson, N. Deshpande, S. Gunduz, U.S. Ozkan, N.A. Brunelli, Utilizing imogolite nanotubes as a tunable catalytic material for the selective isomerization of glucose to fructose, Catal. Today 323 (2018), 69-75.

[195] M. Rochkind, S. Pasternak, Y. Paz, Using dyes for evaluating photocatalytic properties: a critical review, Molecules 20 (2014) 88–110.

[196] E. Poli, J.D. Elliott, L.E. Ratcliff, L. Andrinopoulos, J. Dziedzic, N.D.M. Hine, A.A. Mostofi, C.-K. Skylaris, P.D. Haynes, G. Teobaldi, The potential of imogolite nanotubes as (co-)photocatalysts: a linear-scaling density functional theory study, J. Phys. Condens. Mat. 28 (2016) 074003.

[197] B.M. Rotoli, P. Guidi, B. Bonelli, M. Bernardeschi, M.G. Bianchi, S. Esposito, G. Frenzilli, P. Lucchesi, M. Nigro, V. Scarcelli, M. Tomatis, P.P. Zanello, B. Fubini, O. Bussolati, E. Bergamaschi, Imogolite: an aluminosilicate nanotube endowed with low cytotoxicity and genotoxicity, Chem. Res. Toxicol. 27 (2014) 1142–1154.

[198] W. Liu, P. Chaurand, C. Di Giorgio, M. De Meo, A. Thill, M. Auffan, A. Masion, D. Borschneck, F. Chaspoul, P. Gallice, A. Botta, J.-Y. Bottero, J. Rose, Influence of the length of imogolite-like nanotubes on their cytotoxicity and genotoxicity toward human dermal cells, Chem. Res. Toxicol. 25 (2012) 2513–2522.

[199] S. van den Brule, E. Beckers, P. Chaurand, W. Liu, S. Ibouraadaten, M. Palmai-Pallag, F. Uwambayinema, Y. Yakoub, A. Avellan, C. Levard, V. Haufroid, E. Marbaix, A. Thill, D. Lison, J. Rose, Nanometer-long Ge-imogolite nanotubes cause sustained lung inflammation and fibrosis in rats, Part. Fibre Toxicol. 11 (2014) 67.

[200] J. Geng, K. Kim, J. Zhang, A. Escalada, R. Tunuguntla, L.R. Comolli, F.I. Allen, A.V. Shnyrova, K.R. Cho, D. Munoz, Stochastic transport through carbon nanotubes in lipid bilayers and live cell membranes, Nature 514 (2014) 612.

[201] R.H. Tunuguntla, R.Y. Henley, Y.-C. Yao, T.A. Pham, M. Wanunu, A. Noy, Enhanced water permeability and tunable ion selectivity in subnanometer carbon nanotube porins, Science 357 (2017) 792–796.

[202] V. Bhavanasi, D.Y. Kusuma, P.S. Lee, Polarization orientation, piezoelectricity, and energy harvesting performance of ferroelectric PVDF-TrFE nanotubes synthesized by nanoconfinement, Adv. Energy Mater. 4 (2014) 1400723.

[203] R. Su, Y. Shen, L. Li, D. Zhang, G. Yang, C. Gao, Y. Yang, Silver-modified nanosized ferroelectrics as a novel photocatalyst, Small 11 (2015) 202–207.

[204] F. Sastre, V. Fornés, A. Corma, H. García, Selective, room-temperature transformation of methane to C1 oxygenates by deep UV photolysis over zeolites, J. Am. Chem. Soc. 133 (2011) 17257–17261.